\documentstyle[12pt]{article}

\let\realverbatim=\verbatim
\let\realendverbatim=\endverbatim
\renewcommand\verbatim{\par\addvspace{6pt plus 2pt minus 1pt}\realverbatim}
\renewcommand\endverbatim{\realendverbatim\addvspace{6pt plus 2pt minus 1pt}}

\def\x{{\bf x}}

\def\a{{\alpha}}

\def\e{{\epsilon}}
\def\D{{\cal D}}

\def\half{{1 \over 2}}
\def\ra{{\rangle}}
\def\la{{\langle}}

\def\ih{{i \over \hbar}}

\def\au{\underline \alpha}
\def\p{{\bf p}}

\def\v{{\bf v}}
\def\x{{\bf x}}
\def\X{{\bf X}}
\def\y{{\bf y}}

\input epsf.tex

\begin{document}

\title{The Interpretation of Quantum Cosmology and the Problem of
Time}

\author{J.J.Halliwell \\ \\ Blackett Laboratory \\ Imperial College
\\ London, SW7 2BZ}

\date{}
\maketitle

\begin{abstract}
The development of quantum cosmology, in which Stephen Hawking
played a crucial role, has frequently encountered substantial
conceptual and technical difficulties related to the problem of
time in quantum gravity and to general issues concerning the
foundations of quantum theory. In this contribution to Stephen's
60th Birthday Conference, I describe some recent work in which
the decoherent histories approach to quantum theory is used to
quantize simple cosmological models and perhaps shed some light on
some of these difficulties.
\end{abstract}

\centerline{Preprint Imperial TP/1-02/23}

\section{Introduction}

It is a truly great pleasure to be part of Stephen Hawking's 60th
birthday celebrations. It is now almost twenty years since I
commenced my PhD studies with Stephen, and I recall that exciting
time very fondly. Numerous researchers were just beginning to
address the conceptual and technical difficulties of quantum
cosmology, and the confusion and controversy at that time led to
some very lively debates in which Stephen played a very central
role. The atmosphere of this conference has reminded me very
pleasantly of those early days.

The work described here concerns some generally interesting issues
in quantum cosmology. This is the area I started working on with
Stephen, and its intriguing problems and difficulties have been
the inspiration for much else I have pursued in physics. Quantum
cosmology was born in the 1960's with the work of Bryce DeWitt,
Charles Misner, John Wheeler and others \cite{DeW1,Mis,Whe}, but
it really took off in the early 1980's with Hartle and Hawking's
seminal paper, {\it The Wave Function of the Universe}
\cite{HaHa}. This development was very pertinent since the
appearance of the inflationary universe scenario at about that
time underscored the urgency to acquire an understanding of
cosmological initial conditions. The new ingredients of quantum
cosmology, compared to its older 1960s version, were two-fold.
First, there was the use of Euclidean path integral methods which
had grown out of their very successful application to black holes
physics. Secondly, and perhaps more importantly, there was the
Hartle-Hawking ``no-boundary'' proposal. This gave, perhaps for
the first time, a genuinely quantum gravitational proposal which
implied cosmological initial conditions. In this proposal, wave
functions satisfying the Wheeler-DeWitt equation
\begin{equation}
{\cal H} \Psi [h_{ij}] = 0
\end{equation}
are given by a sum over histories
expression of the form
\begin{equation}
\Psi [h_{ij}] = \int_C {\cal D} g_{\mu
\nu} \exp \left( - I [g_{\mu \nu} ] \right)
\end{equation}
where the sum is over closed four-geometries whose only
boundary is the three-surface on which the three-metric
$h_{ij}$ is specified. (Note also that the functional
integral is taken to be over a {\it complex} contour
$C$ \cite{HalH2}).

There were two questions of particular interest in these early
days. Firstly, what sort of picture can quantum gravity give about
the initial singularity? And secondly, what initial conditions
does the theory imply for classical cosmological solutions?
In simple, somewhat heuristic models, the answer to the first
question is that the initial singularity is replaced by some kind
of quantum tunneling situation. That is, the classical singularity
can (at least in some models) become a classically forbidden
region. The second question can be answered by showing how quantum
cosmology produces, in a quasiclassical limit, an approximate
probability distribution on the space of classical cosmological
solutions.

The most pressing questions in those days were to do with
cosmological issues: Does the no-boundary proposal predict initial
conditions which produce inflationary solutions
and subsequent structure
formation \cite{HaPa,HalHaw}? The interpretation associated with these models was
heuristic \cite{Kuc,Hal1,Hal7,HaPa}. For example, one considered
the Klein-Gordon flux
\begin{equation}
J = i \left( \Psi^* \nabla \Psi - \Psi \nabla \Psi^* \right)
\end{equation}
associated with the wave functions of the WKB type,
\begin{equation}
\Psi = C e^{ i S}
\end{equation}
These heuristic methods appeared to be adequate for the questions
considered at the time.

Now, however, in 2002, far more is known about the foundations
and interpretation of quantum theory. Moreover, there has also
been a considerable amount of activity in the experimental tests
of foundational ideas. As a result of this, the foundations of
quantum theory is now more in the mainstream of physics, when
previously it was regarded as the domain of the philosophers.
Perhaps as a result of this, there has been a secondary wave of
much slower activity in quantum cosmology, that aims to look more
deeply into a whole variety of mathematical and conceptual issues
related to the application of quantum theory to the universe as a
whole. In particular, it is now reasonable to ask, how can the
heuristic ideas used earlier be derived or reconciled with a
properly defined interpretational structure for quantum theory of
models described by a Wheeler-DeWitt equation? For example, in
standard quantum theory, probabilities generally have the form
\begin{equation}
P = {\rm Tr} \left( P \rho \right)
\end{equation}
where
$\rho $ is the density operator, $P$ is a projection operator and
the trace is over a complete set of states (and so assumes a
Hilbert space of states with a suitably defined inner product).
How can the predictions of quantum cosmology be reconciled with a
formula of this type?

Related to these issues is the notorious problem of time
\cite{Kuc,Ish,BuIs}. The Wheeler-DeWitt equation, in simple
minisuperspace models, has the form,
\begin{equation}
H \Psi = \left( - \nabla^2 + U \right) \Psi = 0
\end{equation}
where $\nabla^2$ is a d'Alembertian type operator. The equation is
of the form of a Klein-Gordon equation in a general curved
spacetime background with a space-time dependent mass term. There
is no single variable to play the role of time, nor is there the
possibility of splitting the solutions into positive and negative
frequency. In the usual Klein-Gordon equation one then resorts to
second quantization, but the Wheeler-DeWitt equation is in some
sense already the second quantized theory, so one has to face the
issue of the lack of a time coordinate more squarely.

In this contribution to the conference, I will describe how the
decoherent histories approach may be used to provide a quantization
of simple minisuperspace models, perhaps avoiding some of
the serious difficulties outlined above, and agreeing with
the heuristic methods used earlier on.

I begin by making two simple observations about the Wheeler-DeWitt
equation that are generally relevant to the discussion. The first
is that, at least classical speaking, the constraint equation,
\begin{equation}
H = f^{ab} p_a p_b + V = 0
\end{equation}
is related to reparametrization invariance. (This is the leftover
of four-dimensional diffeomorphism invariance after the
restriction to minisuperspace). In keeping with Dirac's general
ideas about quantizing constrained systems, we therefore look for
{\it observables}-- quantities which commute with the constraint.
There has been a certain amount of debate about this issue in the
context of general relativity, but this will not affect us here
\cite{DeW,Rov3}.

Second, as observed by Barbour \cite{Bar1,Bar2}, by analogy with Mott's
1929 calculation of tracks in a cloud chamber \cite{Mot}, there is a natural
association between the Wheeler-DeWitt equation and the emergence
of classical trajectories. Mott asked why the outgoing spherical
wave associated with alpha decay produced a straight line track
in a cloud chamber. By considering the Hamiltonian of the decaying
atom interacting with ionizing particles in the cloud chamber he
showed that the ionized particles with high probability lie
along a straight line. Interestingly, and as Barbour observed,
Mott actually solved the time-independent Schr\"odinger equation,
\begin{equation}
H \Psi = E \Psi
\end{equation}
rather than the time-dependent one, to obtain the tracks. This is
therefore an interesting analogy for the Wheeler-DeWitt equation
where, in the quasiclassical limit one expects to obtain emergent
classical trajectories. (In fact, a  model of the Wheeler-DeWitt
equation including explicit detectors, and exhibiting these
features has been constructed \cite{Hal2}. Some related
approaches to timeless models are Refs.\cite{Mon,Rov1,Rov2}.)

With these preliminaries out of the way, we now focus on the
following simple question: Suppose we have an $n$ dimensional
configuration space with coordinates $\x = (x_1, x_2, \cdots x_n)$,
and suppose the wave function of the system is in an eigenstate
of the Hamiltonian,
\begin{equation}
H \Psi (x_1, x_2, \cdots x_n ) = E \Psi ( x_1, x_2, \cdots x_n )
\end{equation}
What is the probability of finding the system in a series
of regions of configuration space $\Delta_1, \Delta_2, \cdots
\Delta_N $
{\it without reference to time}?

This question will form the main focus of the rest of this
paper.
We first consider the classical case, and then
the decoherent histories analysis.

\section{The Classical Case}

We begin by considering the classical case which contains many
almost all the key features of the problem.
(We follow the treatment of Ref.\cite{HaTh2} quite closely).
For simplicity we will
concentrate on the case of a single region of configuration space
$\Delta$.

We will consider a classical system described by a
$2n$-dimensional phase space, with coordinates
and momenta $ (\x, \p ) = (x_k, p_k) $, and Hamiltonian
\begin{equation}
H = {\p^2 \over 2 M} + V (\x )
\label{2.1}
\end{equation}
More generally, we are interested in a system for which the
kinetic part of the Hamiltonian has the form $ g^{kj} (\x) p_k
p_j$, where $g^{kj} (\x)$ is an inverse metric of hyperbolic
signature. Most minisuperspace models in quantum cosmology have a
Hamiltonian of this form. However, the focus of this paper is the
timelessness of the system, and the form of the configuration
space metric turns out to be unimportant. So for simplicity, we
will concentrate on the form Eq.(\ref{2.1}).

We assume that there is a classical phase space distribution
function $w (\p, \x ) $, which is normalized according to
\begin{equation}
\int d^n p \ d^n x \ w (\p, \x ) = 1
\label{2.2}
\end{equation}
and obeys the evolution equation
\begin{equation}
{ \partial w \over \partial t } = \sum_k
\left( - { p_k \over M} { \partial w \over \partial x_k}
+ { \partial V \over \partial x_k } { \partial w \over \partial p_k
} \right) = \{ H, w \}
\label{2.3}
\end{equation}
where $\{ \ , \ \}$ denotes the Poisson bracket. The interesting
case is that in which $w$ is the classical analogue of an energy
eigenstate, in which case $ \partial w / \partial t = 0 $, so the
evolution equation is simply
\begin{equation}
\{ H , w \} = 0
\label{2.4}
\end{equation}
It follows that
\begin{equation}
w( \p^{cl} (t), \x^{cl} (t) ) = w ( \p (0), \x (0) )
\label{2.5}
\end{equation}
where $\p^{cl}(t), \x^{cl}(t)$ are the classical solutions with
initial data $\p(0), \x(0)$, so $w$ is constant along the classical
orbits. (The normalization of $w$ then
becomes an issue if the classical orbits are infinite, but we will
return to this in the quantum case discussed below).

Given a set of classical solutions $ ( \p^{cl}(t), \x^{cl}(t) ) $,
and a phase space distribution function $w$, we are interested
in the probability that a classical solution will pass through
a region $\Delta $ of configuration space. We construct
this as follows. First of all we introduce the characteristic
function of the region $\Delta$,
\begin{equation}
f_{\Delta} (\x )= \cases{1,&if $ \x \in \Delta $; \cr
0 &otherwise.}
\label{2.6}
\end{equation}
To see whether the classical trajectory $\x^{cl} (t)$
intersects this region, consider the phase space function
\begin{equation}
A( \x, \p_0, \x_0) =
\int_{-\infty}^{\infty} dt \ \delta^{(n)} ( \x - \x^{cl} (t) )
\label{2.7}
\end{equation}
(In the case of periodic classical orbits, the range of $t$ is
taken to be equal to the period). This function
is positive for points $\x $ on the classical trajectory
labeled by $\p_0, \x_0 $
and zero otherwise.
Hence intersection of the classical
trajectory with the region $\Delta$ means,
\begin{equation}
\int d^n \x \ f_{\Delta} (\x )
\int_{-\infty}^{\infty} dt \ \delta^{(n)} ( \x - \x^{cl} (t) ) > 0
\label{2.8}
\end{equation}
Or equivalently, that
\begin{equation}
\int_{-\infty}^{\infty} dt
\ f_{\Delta} (\x^{cl} (t) ) > 0
\label{2.9}
\end{equation}
This quantity is essentially the amount of parameter time the
trajectory spends in the region $\Delta $.
We may now write down the probability for a classical
trajectory entering the region $\Delta $. It is,
\begin{equation}
p_{\Delta} = \int d^n p_0 d^n x_0 \ w (\p_0, \x_0 )
\ \theta \left(
\int_{-\infty}^{\infty} dt
\ f_{\Delta} (\x^{cl} (t) ) - \e \right)
\label{2.10}
\end{equation}
In this construction, $\e$ is a small positive
number that is eventually sent to zero, and is included
to avoid possible ambiguities in the $\theta$-function
at zero argument. The $\theta$-function
ensures that the phase space integral is
over all initial data whose corresponding classical
trajectories spend a time greater than $\e$ in the
region $\Delta$.

The classical solution $ \x^{cl} (t) $ depends on
some fiducial initial coordinates and momenta,
$ \x_0 $ and $\p_0$, say. In the case of a
free particle, for example,
\begin{equation}
\x^{cl} (t) = \x_0 + {\p_0 t \over M}
\label{2.11}
\end{equation}
The construction is independent of the choice of fiducial initial
points. If we shift $\x_0$, $\p_0$ along the classical
trajectories, the measure, phase space distribution function $w$
and the $\theta$-function are all invariant. Hence the integral over
$\x_0$, $\p_0$ is effectively a sum over classical trajectories.
The shift along the classical trajectories may also be thought of
as a reparametrization, and the quantity (\ref{2.10}) is in fact a
reparametrization-invariant expression of the notion of a
classical trajectory. This means that the probability (\ref{2.10}) has
the form of a phase space overlap of the ``state" with a
reparametrization-invariant operator.

It is useful also to write this result in a different
form, which will be more relevant to the results we get
in the quantum theory case. In the quantum theory,
we generally deal with propagation between fixed points
in configuration space, rather than with phase space point.
Therefore, in the free particle case, consider the change
of variables from $\x_0, \p_0$ to $\x_0, \x_f$, where
\begin{equation}
\x_f = \x_0 + { \p_0 \over M} \tau
\label{2.12}
\end{equation}
Hence $\x_f$ is the position after evolution for
starting from $\x_0$ for parameter time $ \tau$.
The probability then becomes
\begin{equation}
p_{\Delta} = {M \over \tau} \int d^n x_f d^n x_0 \ w (\p_0, \x_0 )
\ \theta \left(
\int_{-\infty}^{\infty} dt
\ f_{\Delta} (\x_0^f (t) ) - \e \right)
\label{2.13}
\end{equation}
where $\p_0 = M (\x_f - \x_0) / \tau $ and
\begin{equation}
\x_0^f (t) = \x_0 + { (\x_f - \x_0) \over \tau} t
\label{2.14}
\end{equation}
The parameter $\tau$ may in fact be scaled out of the whole
expression, hence the probability is independent of it.

\bigskip

\centerline{\it Figure 1}

{\it The rewritten classical probability (\ref{2.13}) in terms of
a sum over initial and final points $\x_0$ and $\x_f$. The
probability for not entering $\Delta$ is a sum over paths as in
case (a). The probability for entering $\Delta$ includes a sum
over classical paths in which $\Delta$ lies between the initial
and final points, as in case (b). But, to agree with the phase
space form of the result (\ref{2.10}), it must also include a sum
over initial and final points for which $\Delta $ does not lie
between them, as in case (c). This figure also applies to the
semiclassical propagator in the quantum case.}

\bigskip

\epsfxsize=.95\textwidth

\epsfbox{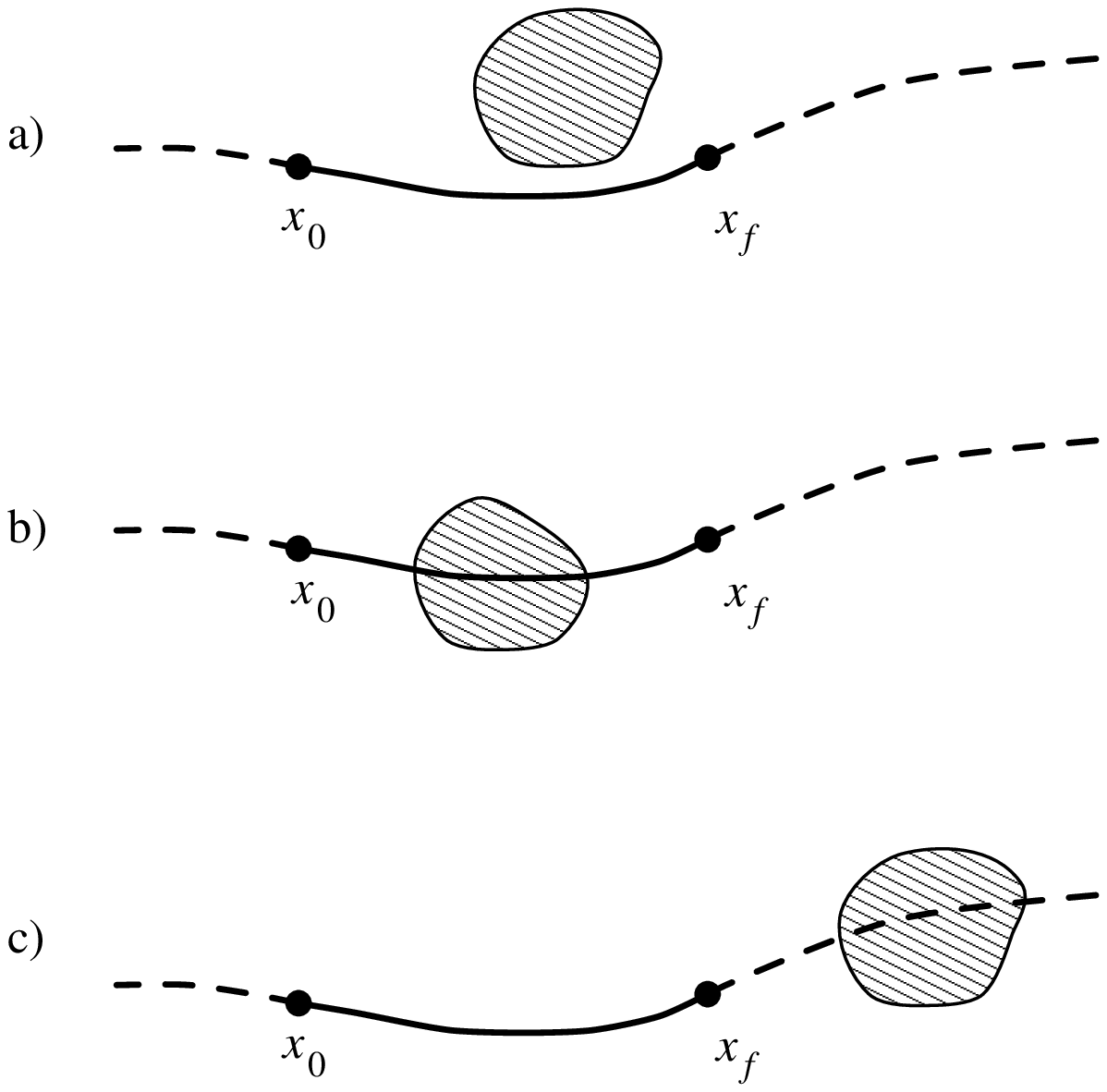}

The result now has the form of an integral over ``initial'' and
``final'' points, analogous to similar results in quantum theory.
The result is again essentially a sum over classical trajectories
with the trajectories now labeled by any pair of points $\x_0$,
$\x_f$ along the trajectories, and is invariant under shifting
$\x_0$ or $\x_f$ along those trajectories. Naively, one might have
thought that the restriction to paths that pass through $\Delta$
is imposed by summing over all finite length classical paths
which intersect $\Delta $ as they go from the ``initial'' point
$\x_0$ to ``final'' point $\x_f$, that is, $\Delta$ lies {\it
between} the initial and final points. This is also what one might
naively expect in the quantum theory version. However, one can see
from the above construction that the correct answer is in fact to
sum over {\it all} classical paths (which can be of infinite
length) passing through $\x_0$ and $\x_f$ that intersect $\Delta$
at {\it any} point along the entire trajectory, even if $\Delta$
does not lie between the two points (see Fig.1). This feature is
related to the reparametrization invariance of the system.

The above point turns out to be quite crucial to what follows
in the rest of this paper, so it is worth saying it in an alternative
form. Loosely speaking, the statement is that only the entire
classical path respects the reparametrization invariance
associated with the constraint equation. A section of the
classical path does not. This may be expressed more precisely
in terms of the function $A(\x, \p_0, \x_0)$ introduced
in Eq.(\ref{2.7}). This function is concentrated on the entire
classical trajectory, and is zero when $\x$ is not on the
trajectory.
It is easy to see that it has vanishing Poisson
bracket with the Hamiltonian $H = H (\p_0, \x_0)$,
since we have
\begin{eqnarray}
\{ H, A (\x, \p_0, \x_0) \}
& = & \int_{-\infty}^{\infty} dt \ \{ H, \delta^{(n)} ( \x - \x^{cl} (t) ) \}
\nonumber \\
&= & - \int_{-\infty}^{\infty} dt \ {d \over dt} \delta^{(n)} ( \x - \x^{cl} (t) )
\nonumber \\
& =  &0
\label{2.15}
\end{eqnarray}
This is the precise sense in which the entire trajectory is
reparametrization invariant, and the phase space function $A$ may
be regarded as an observable -- a quantity which commutes with the
constraint $H$ \cite{Rov3,Mar1}. By way of comparison, consider a
second phase space function similarly defined, but on only a
finite section of trajectory,
\begin{equation}
B( \x, \p_0, \x_0) =
\int_{0}^{\tau} dt \ \delta^{(n)} ( \x - \x^{cl} (t) )
\label{2.16}
\end{equation}
It is easily seen that
\begin{equation}
\{ H, B (\x, \p_0, \x_0) \}
= - \delta (\x - \x^{cl} (\tau) ) + \delta ( \x - \x^{cl} (0) )
\label{2.17}
\end{equation}
Hence $B$ ``almost'' commutes with $H$, failing only
at the end points, and it is in this sense that a finite
section of trajectory does not fully respect reparametrization
invariance.

A third version of the classical result is also useful. It is of
interest to obtain an expression for the probability for
intersecting an $(n-1)$-dimensional surface $\Sigma$. Since the
result (\ref{2.10}) involves the parameter time spent in a finite volume
region $ \Delta $ it does not apply immediately. However, suppose
that the set of trajectories contained in the probability
distribution $w$ intersect the $(n-1)$-dimensional surface
$\Sigma$ only once. Then we may consider a finite volume region
$\Delta$ obtained by thickening $\Sigma$ along the direction of
the classical flow. If this thickening is by a small (positive)
parameter time $\Delta t$, then the quantity appearing in the
$\theta$-function in (\ref{2.10}) is
\begin{eqnarray}
\int dt \int_{\Delta} d^n x \ \delta^{(n)} (\x - \x^{cl} (t))
&=& \Delta t \int dt \int_{\Sigma} d^{n-1} x \ {\bf n} \cdot { d \x^{cl} (t)
\over dt }
\nonumber \\
& & \quad \times \ \delta^{(n)} (\x - \x^{cl} (t))
\nonumber \\
&=& \Delta t \ I [ \Sigma, \x^{cl} (t) ]
\label{2.18}
\end{eqnarray}
where ${\bf n}$ is the normal to $\Sigma$, and we suppose
that the normal is chosen so that ${\bf n} \cdot \ d \x^{cl} /dt $
is positive. The quantity
$I [\Sigma, \x^{cl} (t)] $, in a more general context,
is the intersection number of the
curve $\x^{cl} (t)$ with the surface $\Sigma$, and takes the value
$0$ for no intersections, or $\pm 1$ (depending on whether there is
an even or odd number of intersections). In this case we have
assumed that the trajectories intersect at most once, hence
$ I = 0 $ or $1$. We then have
\begin{equation}
\theta ( \Delta t I  - \e ) = \theta ( I - \e' ) = I
\label{2.19}
\end{equation}
(where $\e = \Delta t \ \e'$)
and the probability for intersecting $\Sigma $ may be written
\begin{equation}
p_{\Sigma} = \int dt \int d^n p_0 d^n x_0 \ w (\p_0, \x_0 ) \ \int_{\Sigma}
d^{n-1} x \ {\bf n} \cdot { d \x^{cl} (t)
\over dt } \ \delta^{(n)} (\x - \x^{cl} (t))
\label{2.20}
\end{equation}
At each $t$, we may perform a change of variables from $\p, \x$
to new variables $\p' = \p^{cl} (t)$, $\x' = \x^{cl} (t)$,
and using Eq.(\ref{2.5}), we obtain the result
\begin{equation}
p_{\Sigma} = {1 \over M} \int dt \int_{\Sigma} d^n p' \ d^{n-1} x' \ {\bf n}
\cdot \p' \ \ w (\p', \x' )
\label{2.21}
\end{equation}
Finally, the integrand is now in fact independent of $t$, so
the $t$ integral leads to an overall factor. (This might be infinite
but is regularized as discussed below). We therefore drop the
$t$ integral.

This result is relevant for the following reason. In the heuristic
``WKB interpretation'' of quantum cosmology, one considers WKB solutions
to the Wheeler-DeWitt equation of the form
\begin{equation}
\Psi = C e^{iS}
\label{2.22}
\end{equation}
It is usually asserted that this corresponds to a set of classical
trajectories with momentum $ \p = \nabla S $, and with a probability
of intersecting a surface $\Sigma$ given in terms of the flux of the
wave function across the surface \cite{Hal1,Hal7,HaPa}.
As we shall show, from the decoherent
histories analysis, the quantum theory gives a probability for
crossing a surface $\Sigma$ proportional to Eq.(\ref{2.21}) with $w$ replaced
by the Wigner function of the quantum theory. The Wigner function of
the WKB wave function is, approximately \cite{Hal7},
\begin{equation}
W(\p, \x ) = \left | C ( \x ) \right|^2 \delta ( \p - \nabla S)
\label{2.23}
\end{equation}
Inserting in Eq.(\ref{2.21}), we therefore obtain, up to overall factors,
the probability distribution,
\begin{equation}
p_{\Sigma} =\int_{\Sigma} d^{n-1} x \ {\bf n} \cdot \nabla S
\ \left| C(\x) \right|^2
\label{2.24}
\end{equation}
We therefore have agreement with the usual heuristic analysis.

\section{The Decoherent Histories Approach to Quantum Theory}

Our aim is to analyze the quantum case using the decoherent
histories approach to quantum theory \cite{DH}.
We first give a very brief review of the formalism.

In non-relativistic quantum mechanics, quantum histories
are represented by so-called class operators $C_{\au}$,
which are given by time-ordered sequences of
projection operators
\begin{equation}
C_{\au} = P_{\a_n} (t_n) \cdots
P_{\a_1} (t_1)
\label{3.1}
\end{equation}
(or more generally, by sums of terms of this form), where
$\au$ denotes the string of alternatives $\a_1, \a_2 \cdots
\a_n$, and $P_{\a_k} (t)$ are projection operators in the
Heisenberg picture. The central object of interest is then the
decoherence functional,
\begin{equation}
D (\au, \au' ) =  {\rm Tr} \left( C_{ \au} \rho
C_{\au'}^{\dag} \right)
\label{3.2}
\end{equation}
Intuitively, the decoherence functional is a measure of the
interference between pairs of histories $\au$, $\au'$. When  its
real part is zero for $\au \ne \au' $, we say that the histories are
consistent and probabilities
\begin{equation}
p (\au ) = D (\au, \au )
\label{3.3}
\end{equation}
obeying the usual probability sum rules may be assigned to them.
Typical physical mechanisms which produce this situation usually
cause both the real and imaginary part of $ D (\au, \au') $
to vanish. This condition is usually called decoherence
of histories, and is related to the existence of so-called
generalized records.

In the non-relativistic case, for histories characterized by
projections onto configuration space, a path integral version of the
decoherence functional is available, and can be very useful. It has
the form,
\begin{equation}
D( \au, \au') = \int_{\au} \D x \ \int_{\au'} \D y
\exp \left( \ih S[x(t)] - \ih S [y(t)] \right)
\rho (x_0, y_0)
\label{3.4}
\end{equation}
where the sum is over pairs of paths $x(t)$, $y(t)$ passing through
the pairs of regions $\au$, $\au'$. This is equivalent to the
form (\ref{3.1}), (\ref{3.2}), when the histories are strings of projections
onto ranges of positions. Eq.(\ref{3.4}) is a useful starting point
for the generalization to timeless theories.

The power of the decoherent histories approach is that it
readily generalizes to a variety of different situations
in which time plays a non-trivial role
\cite{CrHa,Whel,YaT}.
In particular, it may be generalized to the question
of interest here, in which the system is in an energy
eigenstate and we would like to answer questions that
do no refer to time in any way. This generalization requires,
however, specification of the inner product used
to construct the decoherence functional, and a prescription
for the construction of the class operators. We consider
each in turn.

\section{The Induced Inner Product}

For many situations, and especially
for the analogous situation in quantum cosmology, the
Hamiltonian has a continuous spectrum so the energy
eigenstates are not normalizable in the usual inner product,
\begin{equation}
\la \Psi_1 | \Psi_2 \ra = \int d^n x \ \Psi_1^* (\x) \Psi_2 (\x )
\label{4.1}
\end{equation}
A way to deal with this has been developed, and goes by the name
of the induced inner product, or Rieffel induction
\cite{Rie,HaMa}. Consider the eigenvalue equation
\begin{equation}
H | \Psi_{E \lambda} \ra = E | \Psi_{E \lambda} \ra
\label{4.2}
\end{equation}
where $\lambda$ denotes the degeneracy. These eigenstates
will typically satisfy
\begin{equation}
\la \Psi_{E' \lambda'} | \Psi_{E \lambda} \ra =
\delta (E - E') \delta_{\lambda \lambda'}
\label{4.3}
\end{equation}
from which it is clear that the inner product diverges
when $E = E'$. The induced inner product on a set of
eigenstates of fixed $E$ is defined, loosely speaking,
by discarding the $\delta$-function $\delta (E-E')$.
That is, the induced or physical inner product is
then defined by
\begin{equation}
\la \Psi_{E \lambda'} | \Psi_{E \lambda} \ra_{phys} =
\delta_{\lambda \lambda'}
\label{4.4}
\end{equation}
This procedure can be defined quite rigorously, and has been
discussed at some length in Refs.\cite{Rie,HaMa}. We will use it
here to construct the decoherence functional. A simple
prescription for using it in the decoherence functional is to
regularize each propagator and energy eigenstate by using a
different energy for each. The final answer will then involve a
number of $\delta$-functions in energy, as in (\ref{4.3}), which are
simply dropped.

Applied to the Klein-Gordon equation in flat space,
where solutions may be split into positive and negative
frequency parts, $\phi = \phi^+ + \phi^- $, the induced
inner product may be expressed in terms of the Klein-Gordon
inner product (which is negative on the negative frequency
solutions):
\begin{equation}
( \phi, \phi )_I = ( \phi^+, \phi^+ )_{KG} - (\phi^-, \phi^-)_{KG}
\label{4.5}
\end{equation}
This is clearly positive definite. That is, the induced inner
product effectively changes the sign of the negative frequency
solutions to make the overall expression positive definite.
Importantly, the induced inner product does {\it not} require
a split into positive and negative frequency solutions, and
so is generally applicable in quantum cosmological models.

Although the induced inner product has been around for a long
time (in the Klein-Gordon case it was noted by Henneaux and
Teitelboim \cite{HeTe}), it is only comparatively recently that it
made an appearance in quantum cosmology.

\section{The Class Operators}

The most important part of the construction of the decoherence
functional for our timeless model is the construction of the
class operators. On the face of it,
a natural prescription for constructing these is
the following expression:
\begin{eqnarray}
C_{\Delta} (\x_f, \x_0) &=&
\int_{-\infty}^{\infty}
d \tau \ e^{- i E \tau} \ \int \D \x (t) \exp \left( i S [\x(t)] \right)
\nonumber \\
& \times & \ \theta \left(
\int_0^{\tau} dt
\ f_{\Delta} (\x (t) ) - \e \right)
\label{5.1}
\end{eqnarray}
This expression consists of a functional integral between
the initial and final points in fixed parameter time
$\tau$ over paths constrained to enter the region $\Delta$
(enforced by the $\theta$-function as in the classical
case), followed by a sum over all possible parameter times
$\tau$ \cite{Har3}. In keeping with the general ideas of
Dirac quantization, we expect that, in order that the
reparametrization invariance of the theory is fully respected,
a properly constructed class operator should be annihilated
by the constraint,
\begin{equation}
H C_{\Delta} = 0
\label{5.2}
\end{equation}
It is straightforward to show that this is indeed the case
in the limit that $\Delta$ becomes the entire configuration
space. However, as shown in Ref.\cite{HaTh2}, this is
not in fact the case when the region $\Delta$ is finite:
one obtains $\delta$-functions on the boundary of
$\Delta$ on the right-hand side of Eq.(\ref{5.2}).

This is a serious difficulty since it means that
reparametrization invariance is in some sense
violated. It is due to the fact that, as the end-points
move from inside $\Delta$ to outside, the
class operator changes discontinuously. This in turn is
related to the fact that a projection operator onto
the region $\Delta$ does not commute with the constraint
equation.

Because of this difficulty, it is necessary to replace
the class operator $C_{\Delta}$ with a modified
class operator $C_{\Delta}'$ which is as much as possible,
defined by a sum over paths passing through $\Delta$
but satisfies the constraint equation everywhere \cite{HaMa}.
At present, there does not appear to be a universally
agreed way to do this, but some suggestions and discussion
of this point were given in Refs.\cite{HaMa,HaTh1,HaTh2}.

Fortunately, in the semiclassical approximation, it
seems clear how to construct the modified propagator.
It is
\begin{equation}
C_{\Delta}' (\x_f, \x_0) = \theta \left( \int_{-\infty}^{\infty} dt
\ f_{\Delta} (\x_0^f (t) ) -\e \right)
\ P(\x_f, \x_0) \ e^{ i A(\x_f, \x_0)}
\label{5.3}
\end{equation}
Here $P e^{iA}$ is the usual unrestricted semiclassical
propagator, so $A(\x_f, \x_0)$ is the classical action
between initial and final points,
and $P$ is a prefactor.
The $\theta$-function here is the same as in the (rewritten) classical
case Eq.(\ref{2.13}) in terms of ``initial'' and ``final'' points, where
$\x_0^f (t)$ denotes that classical
path from $\x_0$ to $\x_f$. (This is exactly as in the classical
case depicted in Fig.1). Note also that
\begin{equation}
\nabla A \cdot \nabla
\ \ \theta \left( \int_{-\infty}^{\infty} dt
\ f_{\Delta} (\x_0^f (t) ) -\e \right) = 0
\label{5.4}
\end{equation}
as may be shown by shifting the $t$ integration.
It follows that the modified class operator is a semiclassical
solution
to the constraint equation, as required.

It is important that
$t$ is integrated over an infinite range in the quantity inside
the $\theta$-function, otherwise the modified class operator would not
in fact satisfy the constraint. Recall that the originally
defined class operator Eq.(\ref{5.1}) contained a similar $\theta$-function,
with a finite range of time integration, which one might have been
tempted to use in the semiclassical approximation, but this class operator
does not in fact satisfy the constraint.

Hence we see that the
difference between the modified and original class operators
in the semiclassical approximation is the difference between using
the entire classical trajectory or using finite segments of it
in the $\theta$-functions. We also see that these modified
class operators are the correct ones to use in order to
be consistent with the discussion of the classical case
and Eq.(\ref{2.13}). There, we saw that it is appropriate to
sum over classical paths intersecting $\Delta $
even if $\Delta$ does not lie
on the segment of classical trajectory {\it between}
$\x_0$ and $\x_f$.
This feature therefore appears to be
necessary for the particular type of
reparametrization invariance used here.
Only the entire trajectory is reparametrization-invariant
notion. A finite section of trajectory is not.
(See Ref.\cite{HalH} for a further discussion of
reparametrizations in this sort of context).

\section{Decoherence and the Decoherence Functional}

In terms of the class operators $C_{\a}$ and an ``initial''
state $\Psi (\x)$ (an eigenstate of $H$),
the decoherence functional is
\begin{equation}
D (\a, \a') = \int d^n x_f d^n x_0 d^n y_0
\ C_{\a} (\x_f, \x_0 ) C_{\a'} (\x_f, \y_0)
\ \Psi (\x_0) \Psi^* (\y_0)
\label{6.1}
\end{equation}
Here, $\a$ denotes the two sets of histories that either
pass through $\Delta $ or not. The decoherence functional will generally
not be diagonal except for special initial states, so a decoherence
mechanism is necessary.

The details of the decoherence mechanism are somewhat lengthy and
may be found in Ref.\cite{HaTh2}. Briefly, one may couple the
system to a generic environment consisting of large number of
degrees of freedom, and then trace out the environmental
coordinates. We do not address the more physical question of what
the environment actually is for minisuperspace models, but it is
sufficient to observe that there are very many fields that are
unobserved in cosmological models, and so may serve as an
environment. The overall result is that,
in the semiclassical approximation, the decoherence
functional Eq.(\ref{6.1}) becomes
\begin{eqnarray}
D (\a, \a') & = & \int d^n x_f d^n x_0 d^n y_0 \ C_{\a} (\x_f,
\x_0 ) C_{\a'} (\x_f, \y_0)
\nonumber \\
\ & \times &  F (\x_f, \x_0, \y_0 )
\ \Psi (\x_0) \Psi^* (\y_0)
\label{6.2}
\end{eqnarray}
where $ F (\x_f, \x_0, \y_0) $ is the influence functional
in the semiclassical approximation. As shown in Ref.\cite{HaTh2},
it typically has the form,
\begin{equation}
F  = \exp \left(  i v^a \Gamma_a - \half v^a v^b \sigma_{ab} \right)
\label{6.3}
\end{equation}
where $\v = \x - \y $, and $\Gamma_a$ and $\sigma_{ab}$ are real
coefficients (and depend only on $\x + \y$,
and $\sigma_{ab}$ is a non-negative matrix.
The form of the influence functional broadly indicates that
interference between different values of $\x$ (or more generally,
different semiclassical trajectories) is suppressed, as expected.
(The matrix $\sigma_{ab}$ may develop a zero eigenvalue, but
this corresponds to the fact that interference between points
along the same trajectory is not suppressed, {\it i.e.}, that
the influence functional is sensitive only to different
trajectories, reflecting the underlying reparametrization
invariance).

Given decoherence, we may therefore compute
the probability for passing through the region $\Delta$.
Introducing the variable $\X_0 = \half (\x + \y)$, and introducing
the Wigner function,
\begin{equation}
W(\p,\X) = { 1 \over (2 \pi)^n } \int d^n v \ e^{- i\p \cdot \v} \
\rho( \X + \half \v, \X - \half \v)
\label{6.4}
\end{equation}
it is readily shown that the probability is
\begin{equation}
p_{\Delta} = \int d^n p_0 d^n X_0
\ \theta \left(
\int_{-\infty}^{\infty} dt \ f_{\Delta} (\X^{cl} (t) ) -\e \right)
\tilde W (\p_0, \X_0)
\label{6.5}
\end{equation}
Here, $\X^{cl} (t)$ is the classical path with initial data
$\X_0, \p_0$, and
we have defined the smeared Wigner function
\begin{eqnarray}
\tilde W (\p_0 , \X_0)
& =&  \int d^n p
\ \exp \left( - \half ( \p_0 - \p - \Gamma)^a
\sigma_{ab}^{-1} ( \p_0 - \p - \Gamma)^b
\right)
\nonumber \\
& & \quad \times \ W (\p, \X_0)
\label{6.6}
\end{eqnarray}
Eq.(\ref{6.5}) is the main result. It shows that the decoherent
histories analysis very nearly produces the classical result
Eq.(\ref{2.10}), except that instead of a classical phase
space probability distribution, we get a smeared Wigner function.
We also see that the coefficient $\Gamma_a$ from the influence
functional produces a small modification in the classical
equations of motion, and that there are fluctuations about
the classical motion governed by $\sigma_{ab}^{-1}$.
(A possible zero eigenvalue for $\sigma_{ab}^{-1}$ simply
means that the exponential in Eq.(\ref{6.6}) becomes a $\delta$-function
in one particular direction, so there are no fluctuations in that
direction).

\section{Summary and Discussion}

The present work shows that it is possible to carry out a
decoherent histories quantization of simple quantum cosmological
models. This quantization method reproduces, approximately,
heuristic methods based on essentially classical ideas. These
models have the key property of not possessing a time parameter,
but this does not appear to present an insurmountable obstruction
to the application of the approach.

Central to both the classical and quantum problems is the notion
of an entire trajectory. At the classical level it appears to be
the appropriate reparametrization-invariant notion for the
construction of interesting probabilities. At the quantum level,
the decoherent histories approach appears to handle the problem in
a natural way, perhaps because it readily incorporates the notion
of trajectory.

This approach to quantizing cosmological models in this way is
certainly only a first bite at the problem, and a list of further
topics and related issues is described at some length in
Ref.\cite{HaTh2}. Whilst it is gratifying that quantum
cosmological models may be successfully quantized using the
decoherent histories approach, what has been particularly
interesting for me is the observation that this area appears to be
a point of confluence for a number of different fields. As
indicated at the beginning of this article, for example, quantum
cosmology puts severe pressure on the foundations of standard
quantum theory. There has also been an interesting interplay going
on between quantum cosmology and the question of time in standard,
non-relativistic quantum mechanics \cite{Time}, and this is likely
to prove fruitful for quantum cosmology. In the long run, quantum
cosmology may or may not be able to provide useful, testable
cosmological predictions. But what is pretty certain is that work
in this area is likely to remain a source of stimulus for a long
time to come.

\section{Acknowledgements}

I am very grateful to Gary Gibbons, Paul Shellard and Stuart
Rankin for organizing such a wonderful meeting. Particular
gratitude goes to Stephen Hawking for giving us something
to have a meeting about

%\vspace{20pt}
%\hrule width 2in
%\vspace{5pt}
%\noindent \LaTeXe\ \textsc{cmmp} guide v1.02


\begin{thebibliography}{4}\label{bib}




\bibitem{Bar1} J.Barbour, \textit{Class.Quant.Grav.} {\bf 11}, 2853 (1994).
% The timelessness of quantum gravity. I. The evidence
% the classical theory.

\bibitem{Bar2} J.Barbour, \textit{Class.Quant.Grav.} {\bf 11}, 2875 (1994).
% The timelessness of quantum gravity. II. The appearance of
% dynamics in static configurations.

%\bibitem{Bar3} J.Barbour, {\it The End of Time: The Next Revolution
%in our Understanding of the Universe} (Weidenfeld and Nicholson,
%1999).

\bibitem{BuIs} J.Butterfield and C.J.Isham, gr-qc/9901024.
%{\it On the emergence of time in quantum gravity}.

%\bibitem{CoHe} K.Hepp, \textit{ Helv.Phys.Acta} {\bf 45}, 237 (1972).
% More recent developments?

\bibitem{CrHa} D.Craig and J.B.Hartle, preprint UCSBTH-94-47 (1998).
% Generalized quantum theory of Bianchi IX Cosmologies.

\bibitem{DeW1} B.S.DeWitt, \textit{Phys.Rev.} {\bf 160}, 1113 (1967).



\bibitem{DeW} B.DeWitt, in {\it Gravitation: An Introduction to
Current Research}, edited by L.Witten (John WIley and Sons, New
York, 1962).
% The quantization of geometry.

\bibitem{DH} M.Gell-Mann and J.B.Hartle, in {\it Complexity,
Entropy and the Physics of Information, SFI Studies in the
Sciences of Complexity}, Vol. VIII, W. Zurek (ed.) (Addison
Wesley, Reading, 1990); and in {\it Proceedings of the Third
International Symposium on the Foundations of Quantum Mechanics in
the Light of New Technology}, S. Kobayashi, H. Ezawa, Y. Murayama
and S. Nomura (eds.) (Physical Society of Japan, Tokyo, 1990);
%{\it Quantum Mechanics in the Light of Quantum Cosmology.}
\textit{Phys.Rev.} {\bf D47}, 3345 (1993); R.B.Griffiths, \textit{
J.Stat.Phys.} {\bf 36}, 219 (1984); \textit{ Phys.Rev.Lett.} {\bf
70}, 2201 (1993); \textit{ Am.J.Phys.} {\bf 55}, 11 (1987);
R.Omn\`es, \textit{ J.Stat.Phys.} {\bf 53}, 893 (1988); {\bf 53},
933 (1988); {\bf 53}, 957 (1988); {\bf 57}, 357 (1989); {\bf 62},
841 (1991); \textit{ Ann.Phys.} {\bf 201}, 354 (1990); \textit{
Rev.Mod.Phys.} {\bf 64}, 339 (1992); J.B.Hartle, in {\it Quantum
Cosmology and Baby Universes}, S. Coleman, J. Hartle, T. Piran and
S. Weinberg (eds.) (World Scientific, Singapore, 1991);
J.J.Halliwell, in {\it Fundamental Problems in Quantum Theory},
edited by D.Greenberger and A.Zeilinger, Annals of the New York
Academy of Sciences, Vol 775, 726 (1994). For further developments
in the decoherent histories approach, particularly adpated to the
problem of spacetime coarse grainings, see C. Isham,
\textit{J.Math.Phys.} {\bf 23}, 2157 (1994);
%{\it Quantum Logic and the Histories Approach to Quantum Theory.}
C. Isham and N. Linden, \textit{J.Math.Phys.} {\bf 35}, 5452
(1994); {\bf 36}, 5392 (1995).

%\bibitem{Gar} C.W.Gardiner, {\it Quantum Noise} (Springer-Verlag,
%Berlin, 1991).

%\bibitem{Gol} H.Goldstein, {\it Classical Mechanics}
%(Addison-Wesley, Reading MA, 1980).

\bibitem{Hal1} J.J.Halliwell, in, {\it Proceedings of the 13th
International Conference on General Relativity and Gravitation},
edited by R.J.Gleiser, C.N.Kozameh, O.M.Moreschi (IOP Publishers,
Bristol,1992). (Also available as the e-print gr-qc/9208001).
% The Interpretation of Quantum Cosmological Models


\bibitem{Hal2} J.J.Halliwell, \textit{Phys.Rev.} {\bf D64}, 044008 (2001).
%``Trajectories for the Wave Function of the Universe from a
%Simple Detector Model'', % gr-qc/0008046.


%\bibitem{Hal2} J.J.Halliwell, \textit{Phys.Rev.} {\bf D48}, 4785 (1993).
% Quantum-mechanical histories and the uncertainty principle.
% II. Fluctuations about classical predictability.

%\bibitem{Hal3} J.J.Halliwell, \textit{Phys.Rev.} {\bf D38}, 2468 (1988).
% Derivation of the Wheeler-DeWitt equation
% from a Path Integral for Minisuperspace Models.

%\bibitem{Hal4} This detector model was used in a simple
%non-relativistic context by J.J.Halliwell, \textit{Phys.Rev.} {\bf
%D60}, 105031 (1999).
% ``Somewhere in the Universe: Where is the Information Stored when
% Histories Decohere?'', quant-ph/9902008, Imperial preprint
% TP/98-99/29 (1999).
%Some subsequent developments of the Coleman-Hepp model are
%H.Nakazato and S.Pascazio, \textit{Phys.Rev.} {\bf 70}, 1 (1993);
%\textit{Phys.Rev.} {\bf A48}, 1066 (1993); R.Blasi, S.Pascazio,
%S.Takagi, \textit{Phys.Rev.} {\bf A250}, 230 (1998).

%\bibitem{Hal5} J.J.Halliwell, \textit{Prog.Theor.Phys.} {\bf 102}, 707 (1999).


\bibitem{Hal7} J.J.Halliwell, \textit{Phys.Rev.} {\bf D36}, 3626 (1987).
%``Correlations in the Wave Function of the Universe",

\bibitem{HalH2} J.J.Halliwell and J.B.Hartle,
\textit{Phys.Rev.} {\bf D41}, 1815 (1990).
% ``Integration Contours for the No-Boundary Wave
% Function of the Universe"



\bibitem{HalH} J.J.Halliwell and J.B.Hartle, \textit{Phys.Rev.} {\bf D43},
1170 (1991).
% ``Wave Functions Constructed from an Invariant
% Sum-Over-Histories Satisfy Constraints",


\bibitem{HalHaw} J.J.Halliwell and S.W.Hawking,
\textit{Phys.Rev.} {\bf D31}, 1777 (1985).
% Origin of Structure in the Universe



%\bibitem{HaOr} J.J.Halliwell and M.E.Ortiz, \textit{Phys.Rev.} {\bf D48}, 748 (1993).

\bibitem{HaTh1} J.J.Halliwell and J.Thorwart, \textit{Phys.Rev.} {\bf
D64}, 124018 (2001)
% Decoherent histories analysis of the relativistic particle

\bibitem{HaTh2} J.J.Halliwell and J.Thorwart,  gr-qc/0201070,
accepted for publication in Physical Review D (2002).
% Life in an Energy Eigenstate: Decoherent Histories Analysis of a
% Model Timeless Universe




%\bibitem{Har1} J.B.Hartle, \textit{Phys.Rev.} {\bf D37}, 2818 (1988).
% Quantum kinematics of spacetime. I. Non-relativistic theory.

%\bibitem{Har2} J.B.Hartle, \textit{Phys.Rev.} {\bf D38}, 2985 (1988).
% Quantum kinematics of spacetime. II. A model quantum cosmology
% with real clocks.

\bibitem{Har3} J.B.Hartle, in {\it Proceedings of the 1992 Les Houches
School, Gravity and its Quantizations}, edited by B.Julia and
J.Zinn-Justin (Elsevier Science B.V. 1995)
% Spacetime quantum mechanics and the quantum mechanics of
% spacetime

\bibitem{HaMa} J.B.Hartle and D.Marolf,  \textit{Phys.Rev.}
{\bf D56}, 6247 (1997).
% Comparing Formulations of Generalized Quantum Mechanics for
% Reparametrization-Invariant Systems


\bibitem{HaHa} J.B.Hartle and S.W.Hawking, \textit{Phys.Rev.} {\bf 28}, 2960
(1983).



\bibitem{HaPa} S.W.Hawking and D.N.Page, \textit{Nucl.Phys.} {\bf B264},
185 (1986);
%Operator ordering and the flatness of the universe.
{\bf B298}, 789 (1988).
%How probable is inflation?


%\bibitem{Haw} S.W.Hawking, \textit{Phys.Rev.} {\bf D32}, 2489 (1985).
% The arrow of time in cosmology.

\bibitem{HeTe} M.Henneaux and C.Teitelboim, \textit{Ann.Phys.(N.Y.}
{\bf 143} 127 (1982).

\bibitem{Ish} C.J.Isham, gr-qc/9210011.
%{\it Canonical quantum gravity and the problem of time}.


\bibitem{KaNa} Y.Kazama and R.Nakayama, \textit{Phys.Rev.} {\bf 32}, 2500 (1985).
% Wave packet in quantum cosmology.

\bibitem{Kie} C.Kiefer, \textit{Phys.Rev.} {\bf D38}, 1761 (1988).
% Wave packets in minisuperspace.

%\bibitem{Kla} J.Klauder, Ann. Phys. (NY) {\bf 254}, 419 (1997),
%(quant-ph/9604033) ; Nucl.Phys. {\bf B547},397, 1999,
%(hep-th/9901010); hep-th/0003297.


\bibitem{Kuc} K.Kuchar, in {\it Conceptual Problems of Quantum
Gravity}, edited by A.Ashtekar and J.Stachel (Boston, Birkhauser,
1991); and in {\it Proceedings of the 4th Canadian Conference on
General Relativty and Relativistic Astrophysics}, edited by
G.Kunstatter, D.E.Vincent and J.G.Williams (World Scientific, New
Jersey, 1992). See also the e-print gr-qc/9304012, {\it Canonical
quantum gravity}.


\bibitem{Mar1} D.Marolf, \textit{Class.Quant.Grav.} {\bf 12}, 1199 (1995).
% Quantum observables and recollapsing dynamics

\bibitem{Mar2} D.Marolf, \textit{Phys.Rev.} {\bf D53}, 6979(1996);
% Path Integrals and Instantons in Quantum Gravity
\textit{Class.Quant.Grav.} {\bf 12}, 2469 (1995);
% Almost Ideal Clocks in Quantum Cosmology: A Brief Derivation of Time
\textit{Class.Quant.Grav.}{\bf 12}, 1441 (1995).
% Observables and a Hilbert Space for Bianchi IX


\bibitem{Mis} C.W.Misner, in \textit{Magic Without Magic:
John Archibald Wheeler, a Collection of Essays in Honor of his
60th Birthday}, edited by J.Klauder (Freeman, San Francisco,1972).

\bibitem{Mon} M.Montesinos, C.Rovelli and T.Thiemann,
%`SL(2,R) model with two Hamiltonian constraints,'
\textit{Phys.Rev.} {\bf  D60}, 044009 (1999);
M.Montesinos,
%`Relational evolution of the degrees of freedom
% of generally covariant quantum theories'
\textit{Gen.Rel.Grav.} {\bf 33}, 1 (2001);
M.Montesinos and C.Rovelli,
% `Statistical mechanics of generally covariant quantum theories:
% a Bolztmann-like approach'
\textit{Class. Quantum Grav.} {\bf 18}, 555 (2001).



\bibitem{Mot} N.F.Mott, \textit{ Proc.Roy.Soc} {\bf A124}, 375 (1929),
% The wave mechanics of alpha-ray tracks.
reprinted in {\it Quantum Theory and Measurement}, edited by
J.Wheeler and W.Zurek (Princeton University Press, Princeton, New
Jersey, 1983). For further discussions of the Mott calculation see
also, J.S.Bell, {\it Speakable and Unspeakable in Quantum
Mechanics} (Cambridge University Press, Cambridge, 1987);
A.A.Broyles, \textit{Phys.Rev.} {A48}, 1055 (1993); M.Castagnino
and R.Laura, gr-qc/0006012.


%\bibitem{PDX} A.Auerbach and S.Kivelson, \textit{Nucl.Phys.} {\bf B257}, 799 (1985).
%The path decomposition expansion and multidimensional tunneling

\bibitem{Rie} A.Ashtekar, J.Lewandowski, D.Marolf, J.Mourao and
T.Thiemann, \textit{ J.Math.Phys.} {\bf 36}, 6456 (1995);
% Quantization of diffeomorphism invariant theorie of connections with
% local degrees of freedom.
A.Higuchi, \textit{Class.Quant.Grav.} {\bf 8}, 1983 (1991).
% Quantum linearization instabilities of de Sitter spacetime 2.
D.Giulini and D.Marolf, \textit{Class.Quant.Grav.} {\bf 16}, 2489
(1999);
% A Uniqueness Theorem for Constraint Quantization
\textit{Class.Quant.Grav.} {\bf 16}, 2479 (1999).
% On the Generality of Refined Algebraic Quantization
F.Embacher, \textit{ Hadronic J.} {\bf 21}, 337 (1998);
% Handwaving refined algebraic quantization.
N.Landsmann, \textit{ J.Geom.Phys.} {\bf 15}, 285 (1995);
% Rieffel induction as generalized quantum Marsden-Weinstein
% reduction
D.Marolf, gr-qc/0011112.
% Group Averaging and Refined Algebraic Quantization: Where are we now?

\bibitem{Rov1} C.Rovelli, \textit{Phys.Rev.} {\bf 42}, 2638 (1990).
% Quantum mechanics without time: A model.

\bibitem{Rov2} C.Rovelli, \textit{Phys.Rev.} {\bf 43}, 442 (1991).
% Time in quantum gravity: An hypothesis.

\bibitem{Rov3} C.Rovelli, \textit{Class.Quant.Grav.} {\bf 8}, 297 (1991);
% What is observable in classical and quantum gravity
{\bf 8}, 317 (1991).
% Quantum references systems.

%\bibitem{Sch} L.Schulman, {\it Techniques and Applications of Path
%Integrals} (Wiley, New York, 1981).

%\bibitem{Tei} C.Teitelboim,
%\textit{Phys.Rev.} {\bf D25}, 3159 (1983); {\bf 28}, 297 (1983);
%{\bf 28}, 310 (1983).

\bibitem{Time} An extensive list of references to the issue of
time in quantum mechanics may be found in \cite{HaTh2}. See also
the forthcoming volume, {\it Time in Quantum Mechanics}, edited by
J.G.Muga, R.Sala Mayato and I.L.Egususquiza (to appear in 2002).



\bibitem{Whe} J.A.Wheeler, in \textit{Batelles Rencontres}, eds.C.DeWitt and
J.A.Wheeler (Benjamin, New York, 1968)

\bibitem{Whel} J.Whelan, \textit{Phys.Rev.} {\bf D50}, 6344
(1994).
% Spacetime alternatives in relativistic particle motion


\bibitem{YaT} N.Yamada and S.Takagi, \textit{ Prog.Theor.Phys.}
{\bf 85}, 985 (1991); {\bf 86}, 599 (1991); {\bf 87}, 77 (1992);
N. Yamada, \textit{ Sci. Rep. T\^ohoku Uni., Series 8}, {\bf 12},
177 (1992); \textit{Phys.Rev.} {\bf A54}, 182 (1996);
J.J.Halliwell and E.Zafiris, \textit{Phys.Rev.} {\bf D57},
3351-3364 (1998); J.B.Hartle, \textit{Phys.Rev.} {\bf D44}, 3173
(1991);
%{\it Spacetime Coarse-Grainings in Non-Relativistic Quantum Mechanics.}
R.J.Micanek and J.B.Hartle, \textit{Phys.Rev.} {\bf A54}, 3795
(1996).
% Nearly instantaneous alternatives in quantum mechanics.

%\bibitem{Zeh} H.D.Zeh, {\it The Physical Basis of the Direction of
%Time}, third edition (Springer-Verlag, 1999) (and the associated
%webpage www.time-direction.de); \textit{ Phys.Lett.} {\bf A116}, 9
%(1986); {\bf A126}, 311 (1988); C.Kiefer and H.Zeh, \textit{
%Phys.Rev.} {\bf D51}, 4145 (1995).

\end{thebibliography}
\end{document}